\documentclass[aps,pra,preprint,eqsecnum,draft,showpacs,showkeys]{revtex4}
\begin{document}
\preprint{}

\title{ALGORITHM OF REDUCTION}
\author{N. K.  Solovarov
}

\affiliation{Zavoisky Physical-Technical Institute of the Russian Academy
of Sciences, Sibirskii trakt 10/7, 420029 Kazan, Russia\\
Kazan State Pedagogical University, ul. Mezhlauk 1, 420021 Kazan, Russia}
\email{solovar@kfti.knc.ru}

\date{\today}
\begin{abstract}
We show that the von Neumann's algorithm of reduction (i.e.
the algorithm of calculating the density matrix of the observable
subsystem from the density matrix of the closed quantum system)
corresponds to the special approximation at which the unobservable
 subsystem is supposed to be in the steady
state of minimum information
(infinite temperature) .
We formulate the generalized  algorithm of reduction that includes
as limiting cases the von Neumann's reduction and
the self-congruent correlated reduction most corresponding
to the quantum nondemolition measurement.\\
We demonstrate the correlation in dynamics of subsystems
with exactly soluble models of quantum optics: 1) about the dynamics of
a pair of interacting two-level atoms, and 2) about the dynamics of a
two-level atom interacting with a single-mode resonant field.
\end{abstract}

\bigskip


\maketitle

\section{INTRODUCTION}
\label{sec1}

The old question of the quantum theory on the reduction of a quantum
system and on the decoherence of an observable subsystem
(transition from a pure state to a mixed state)
due to its interaction with an unobservable subsystem and the measurement is
formulated anew in quantum optics \cite{klysh,men}.
 This process is initiated by problems of
 quantum computing  \cite{kilin}, by hope to clarify the quantum nature of the
 irreversibility \cite{prigog,men} and, mainly, by the new experimental potentials
to check consequences of  axioms of quantum mechanics accepted previously.
 The qualitatively new experiments  allow one to observe
dynamics of subsystems of simple quantum systems (e.g., the dynamics of
a pair of correlated photons emitted in a nonlinear process, or the
dynamics of a single two-level
atom in the course of its interaction with the resonant field of extremely
small intensity \cite{klysh,men,kilin,haroche}).
The mathematical operation of reduction defines the rule of calculating the
density operator of an observable subsystem from the known density operator
of a closed quantum system. By the von Neumann's postulate, this rule consists in
calculating the partial trace of the density matrix of a closed quantum
system over the unobservable subsystem \cite{von,dirac,fano,blum}.
Though this nonunitary
operation has been universally accepted
 to account for the influence of the classical measuring device
 on the result of measurement,
 to realize the transition from the closed system to the open one, and
  to introduce the irreversibility into the
 dynamics of an observable subsystem \cite{prigog,von,dirac,blum,men},
the algorithm of
 accounting for the interaction of the closed quantum system with
a measuring device remains in so doing reduction implicit. Moreover, when
calculating
the mean values of observables of subsystems, the realization of
 reduction is implicit ("disguised") by itself.

The initial point of the our treatment lies in the identity of the mathematical
formulation of the reduction (other than the interpretation!) with the
formulation of the quantum relaxation theory problem \cite{fano,blum}. In
both cases the closed composite quantum system is considered. Consequently
the unitary time evolution of the system is described by the density operator
$\rho(t)$, which may be approximately calculated (basically in general case).
In both cases the dynamics, i.e. the density operator, of one subsystem
$\hat\rho_{\alpha}(t)$ of the system should be calculated. This extraction
of the one subsystem implies implicitly the separation of the system into
two parts: the observable subsystem $\alpha$ and the unobservable subsystem
$\beta$. In general case the subsystems of the arbitrary closed system are
equal in rights. Let us suppose the dynamics of the subsystem $\beta$ has
described by some density operator $\hat\rho_{\beta}(t)$. The exact
representation of the density operator of the closed system through the density
operators of subsystems is known for the noninteracting subsystems only:
$\hat\rho(t)=\hat\rho_{\alpha}(t)\otimes\hat\rho_{\beta}(t)$ \cite{fano,blum}.
The considered problem consist in calculating of the density operator of
one subsystem in the general case of interacting or interacted previously
(entangled) subsystems.

In the quantum relaxation theory the "unobservable" subsystem is believed
to be the thermostat at a temperature of $T$ from physical reasons.
Consequently its density operator supposed to be known and stationary:
$\hat\rho_{\beta}(T)$. The desired density operator $\hat\rho_{\alpha}(t)$ next
is calculated to some approximation from the approximate equation
$\hat\rho(t)\approx\hat\rho_{\alpha}(t)\otimes\hat\rho_{\beta}(T)$, where
two operators $\hat\rho(t)$ and $\hat\rho_{\beta}(T)$ are known. In the
reduction problem the state of the unobservable subsystem is believed unknown.
The algorithm was instead postulated to calculate $\hat\rho_{\alpha}(t)$
from $\hat\rho(t)$. The question is arises then: is this postulation equivalent
to some approach on the state of unobservable subsystem? As a matter of fact
this question is identical with the mathematical problem of the quantum
relaxation theory: how the density operator of subsystem $\hat\rho_{\beta}(t)$
should be approximately calculated from the known density operators
$\hat\rho(t),\hat\rho_{\alpha}(t)$?

We apply the method similar to used by von Neumann for demonstrating the
correlation (the one-to-one correspondence) between the physically observables
of the different interacting subsystems of the closed quantum system
\cite{von}. In this way the approximate equation system demonstrating the
mutual relations between the density operators of subsystems and the density
operator of the closed system was deduced. This equation system reveals that
given in some way the density operator of one subsystem, the density operator
of the other subsystem can be calculated approximately.
We prove that the von Neumann's algorithm
of reduction is identical to the approximation when the subsystem $\beta$
is constantly in the state of minimum information $\hat\rho_{\beta min}$
(identical with the state of infinite temperature \cite{von,fano}). It should
emphasize this approximation is solely employing when the reduction carring
out, other than when the $\hat\rho(t)$ calculating!

By this means the von Neumann's reduction is mathematically equivalent to
the replacement $\hat\rho(t)\to\hat\rho_{\alpha}(t)\otimes\hat\rho_{\beta
min}$ with exactly defined operators $\hat\rho_{\alpha}(t),\hat\rho_{\beta
min}$. This replacement corresponds on other hand to the changing of the
considered quantum system and to the traditional interpretation of reduction
as the transition from the closed to an open system or the accounting for
the measurement. The state of minimum information (of infinite temperature)
corresponds to the complete absence of the quantum coherence in the subsystem
$\beta$. This implicit approach leads to some decoherence of the observable
subsystem although all the calculations are done in framework of the closed
quantum system.

We get the conclusion that any one of the possible algorithms of reduction
is mathematically equivalent to the replacement
$\hat\rho(t)\to\hat\rho_{\alpha}(t)\otimes\hat\rho_{\beta}(t)$, i.e. to
the approximate representation of the closed quantum system as two
quasy-independent subsystems with the definite dynamics. This raises the
question of whether the accepted algorithm of reduction (i.e. the corresponding
replacement of the density operator) corresponds to the initial assumption of
the closed quantum
system. Or mathematically: how much would this replacement be in error with
the believed to be true $\hat\rho(t)$?

As it is evident from ours consideration, the generally accepted von Neumann's
algorithm of reduction is self-contradictory: on the one hand the closed
quantum system is assumed, and on the other the infinite temperature of the
subsystem is taken implisitly. It is possible to perfect the algorithm of
reduction: for example, to accept some more reasonable state of the
unobservable subsystem (the thermostat, as in the quantum relaxation theory,
or one of eigenstates, as in the theory of quantum nondemolition measurements
of photons \cite{imoto,fucuo}, or any intermediate variant). But in any case
this step corresponds to the changing of the considered system and to the
coming to the new system with the new dynamics, what is customary interpreted
as the coming to the open system.

If the both subsystems of the considered system are the essentially quantum
systems and the approximation of the closed system is believed true, it seems
reasonable to use the principle of the interchangeability of subsystems
\cite{von} to make the reduction. We use the mutual relations between the
density operators $\hat\rho(t),\hat\rho_{\alpha}(t),\hat\rho_{\beta}(t)$
to calculate the correlated density operators $\hat\rho_{\alpha
C}(t),\hat\rho_{\beta C}(t)$ simultaneously by the self-congruent procedure
of the successive approximations. In this way the reduction making the minimum
(the vanishing in the limit) changing of $\hat\rho(t)$ is obtainable. In
other words, the correlated reduction procedure is most consistent with the
initial approximation of the closed quantum system or to the quantum
nondemolition measurement.

The structure of the paper is as follows: in Sec. II we demonstrate
that  the von Neumann's reduction corresponds to the approximation
of the steady state of minimum information of the unobservable subsystem and
 formulate the generalized reduction algorithm as an approximation
 of a given state of the unobservable subsystem of a closed quantum system.
 Sec. III contains the algorithm of correlated reduction that defines the
procedure of calculating self-congruent density operators of subsystems of
a closed quantum system in the limit of the quantum nondemolition
measurement. In Sec. IV we use the correlated reduction technique to
consider three known problems of quantum optics: 1) the separable
representation of the entangled EPR(Einshtein-Podolsky-Rosen) state,
2) the dynamics of a pair of interacting two-level atoms, 3) the dynamics of
a two-level atom interacting with the resonant single-mode electromagnetic
field (the model of cavity quantum electrodynamics).

{\bf Designations and definitions} \cite{von,dirac,fano,blum}:

1. $\hat H =\hat H_{\alpha}+\hat H_{\beta}+\hat H_{\alpha\beta} $ is a
hamiltonian of a closed quantum system,
$\hat H_{\alpha}$ and $\hat H_{\beta} $ are hamiltonians of uncoupled subsystems
$\alpha$ and $\beta$, respectively,
$\hat H_{\alpha\beta}$ is the interaction hamiltonian of subsystems;
 $\hat A$ and $\hat B$ are  operators of observables of subsystems
 $\alpha$ and $\beta$, respectively.

2. $\hat\rho(t) = \hat U(t)\hat\rho(0)\hat U^{\dagger}(t)$ is the solution of
the Schr\"odinger equation determining the dynamics of the system
$\hat H$,
 $\hat U(t)=\exp[-(it/\hbar)\hat H] $ is the operator of evolution.

3. Operators $\hat H_{\alpha}, \hat A$ and $\hat H_{\beta}, \hat B$ are
 determined in spaces of subsystems $\alpha$ and $\beta$, respectively;
 operators $\hat H, \hat H_{\alpha\beta}, \hat\rho(t), \hat U(t) $
are determined in the complete space that is the direct
(exterior $\leftrightarrow\otimes$) product of spaces
$\alpha,\beta;$
 $\hat A'=\hat A\otimes\hat 1_{\beta},\hat B'=\hat 1_{\alpha}\otimes\hat B$
are extended operators of subsystems in the complete space;
 $\hat 1_{\alpha(\beta)}$ are unity operators in spaces
of subsystems,
 $\hat 1_{\alpha}\otimes\hat 1_{\beta}=\hat 1$ is the unity operator
of the complete space;
$ a_{ij}, b_{i'j '}$ are matrix elements of the operators
$\hat A, \hat B$;
$ o_{\{ii'\}\{jj'\}}$ are matrix elements of operators ($\hat O$)
of the closed system, a pair of subscripts $\{ii'\}$ designates basis state
of the closed system.

 4. The von Neumann's algorithm of calculating the reduced density
 operator of the observable subsystem is determined by:
\begin{equation}
\hat\rho_{\alpha N}(t)=Sp_{\beta}\hat\rho(t),\qquad\mbox{or}\qquad
\hat\rho_{\beta N}(t)=Sp_{\alpha}\hat\rho(t),
\label{1}
\end{equation}
where $Sp_{\alpha(\beta)}$  means the partial trace over
the subsystem $\alpha(\beta)$, the subscript $N$ hereinafter designates
 quantities
 calculated in accordance with the von Neumann's algorithm
of reduction.

\section{THE THEOREM ON THE CORRELATION OF SUBSYSTEMS}

{\bf Theorem}

{\it
The reduction $(\ref{1})$ of the closed quantum system $\hat H$
for the observable subsystem $\alpha$ corresponds to the approximation
 at which: {\bf i)} the closed system is represented as
 two uncoupled correlated subsystems, and {\bf ii)} the unobservable
 subsystem $\beta$ is supposed to be in the steady state of
 minimum information.}

 {\bf Proof}.

{\bf i)}. Let us suppose that the dynamics of
 the closed system
 $\hat H$ is approximately represented as the dynamics of two uncoupled
  subsystems. It means \cite{fano,blum} that
 there are such density operators
$\hat\rho_{\alpha C}(t)$ and
 $\hat\rho_{\beta C}(t)$ ($Sp_{\alpha}\hat\rho_{\alpha C}(t)=
 Sp_{\beta}\hat\rho_{\beta C}(t)=1$) of subsystems for which
\begin{equation}
\hat\rho_{\alpha C}(t)\otimes\hat\rho_{\beta C}(t)\approx\hat\rho(t).
\label{2}
\end{equation}
We transform
(\ref{2}) in much the same way as von Neumann did to prove the mutual
correlation of mean values
$\langle\hat A(t)\rangle,\langle\hat B(t)\rangle $
 ( see \cite[Ch.6.2]{von}).  If we multiply (\ref{2}) on both sides by
 $\hat\rho^{'}_{\beta C}(t)=\hat 1_{\alpha}\otimes\hat\rho_{\beta C}(t)$
or $\hat\rho^{'}_{\alpha C}(t)=\hat\rho_{\alpha C}\otimes\hat 1_{\beta}$
and take partial traces over the subsystem $\beta $ or $\alpha $ we obtain
the system of two connected equations \cite{solo,soloann}:
\begin{equation}
\hat\rho_{\alpha C}(t)\approx\frac{Sp_{\beta}\hat\rho(t)
\hat\rho^{'}_{\beta C}(t)}
{Sp_{\alpha\beta}\hat\rho(t)\hat\rho^{'}_{\beta C}(t)},\label{3a}
\end{equation}
\begin{equation}
\hat\rho_{\beta C}(t)\approx\frac{Sp_{\alpha}\hat\rho(t)
\hat\rho^{'}_{\alpha C}(t)}
{Sp_{\alpha\beta}\hat\rho(t)\hat\rho^{'}_{\alpha C}(t)}.\label{3b}
\end{equation}
The right-hand sides of (\ref{3a}),(\ref{3b}) represent the normalized quantum
 averaging of the density operator of the closed system over one of
the subsystems. Each equation  determines the density operator of one
 subsystem through the density operator
of the closed system and density operator of another subsystem.
 From (\ref{2}) and (\ref{3a}),(\ref{3b}), it is evident that density operators
 $\hat\rho_{\alpha C}(t),\hat\rho_{\beta C}(t)$
are correlated, that is why we may call {\bf i)} the approximation
of correlated subsystems (is marked by a subscript {\sl C}).

{\bf ii).} Let us suppose that the unobservable subsystem $\beta $ is in
the steady state of minimum information, i.e. its density operator
equals \cite{von,fano}:
\begin{equation}
\hat\rho_{\beta C}(t)=\hat\rho_{\beta\, min}=\left(1/N_{\beta}\right)
\hat 1_{\beta},\label{4}
\end{equation}
where $N_{\beta}$ is the dimension (number of eigenstates) of
the subsystem $\beta $.
The theorem then follows if we insert (\ref{4}) into the right-hand side of
(\ref{3a}):
\begin{equation}
\hat\rho_{\alpha}(t)\approx\frac{Sp_{\beta}\hat\rho(t)(1/N_{\beta})
\left(\hat 1_{\alpha}\otimes\hat 1_{\beta}\right)}
{Sp_{\alpha\beta}\hat\rho(t)(1/N_{\beta})
\left(\hat 1_{\alpha}\otimes\hat 1_{\beta}\right)}=
Sp_{\beta}\hat\rho(t)=\hat\rho_{\alpha N}(t).\label{5}
 \end{equation}

{\bf Consequence 1}

By (\ref{2})-(\ref{4}), the von Neumann's algorithm of reduction (\ref{1})
is equivalent
to the replacement of the density operator $\hat\rho(t)$ with the reduced
density operator $\hat\rho_{RN}(t)$ according to the rule:
\begin{equation}
\hat\rho(t)\to\hat\rho_{RN}(t)=
\hat\rho_{\alpha N}(t)\otimes\hat\rho_{\beta min}=
\left[Sp_{\beta}\hat\rho(t)\right]\otimes
\left(1/N_{\beta}\right)\hat 1_{\beta}\label{6}.
 \end{equation}The relation (\ref{6}) agrees in form with "
 the basic
equation of irreversibility " or
 "the first assumption" of the quantum relaxation theory
 ( see, e.g. \cite[Ch.7]{blum}):
\begin{equation}
\hat\rho(t)\to
\left[Sp_{\beta}\hat\rho(t)\right]\otimes
\frac{\exp\left[-\left(1/k_{B}T\right)\hat H_{\beta}\right]}
{Sp_{\beta}\exp\left[-\left(1/k_{B}T\right)\hat H_{\beta}\right]},
\label{7}\end{equation}
that redefines $\hat\rho(t)$ in a case, when the unobservable subsystem
is taken to be the thermostat at a temperature of $T $ ($k_{B}$ is
the Boltzmann constant).
It follows
from the comparison of (\ref{6}) with (\ref{7}) that the approximation
{\bf ii)} within the quantum relaxation theory
 corresponds to the representation of the unobservable subsystem
by the thermostat at the infinite temperature, no matter of actual physical
properties of the unobservable subsystem.

{\bf Consequence 2}

{\bf i)}, and {\bf ii)} represent qualitatively distinct approximations.
The first approximation ({\bf i}) represents a closed quantum
 system as two of such quasi-uncoupled correlated subsystems with
 mean values of all observables of the closed system being approximately
 correct.
The second approximation ({\bf ii}) describes the action
 of the measurement
on the state of the observable subsystem and implicitly carries out the
transition from a closed system to an open one.
By Consequence 1, the  model of measurement
(\ref{1})$\equiv$(\ref{6}) used in the von Neumann's reduction may be
considered as the identification of the
classical measuring device with a thermostat of the infinite
temperature and the measurement with the ideal thermal contact of the
unobservable subsystem with this device.
The limited applicability of such model and hence the von Neumann's
 reduction seems physically evident.

In the  form similar to {\bf ii)}, the model of measurement
was included into the algorithm of reduction in the theory of
quantum nondemolition measurements of quantum optics \cite{imoto,fucuo}.
The "probe"
(unobservable) subsystem $\beta$ was postulated to be determined with
certainty in its eigenstate
 $\vert j'\rangle $ at the measurement point of time $t$, i.e.
the probe subsystem density operator is to be presented by a projector
 $\hat\rho_{\beta j'}(t)=\hat P_{\beta j'}\equiv\vert j'\rangle\langle j'
\vert. $
The equation of such "projective" reduction (is marked by the
subscript $P$) was determined by \cite{imoto,fucuo}:
\begin{equation}
\hat\rho(t)\to\hat\rho_{RPj'}(t)=
\hat\rho_{\alpha j'}(t)\otimes\hat P_{\beta j'}=
\frac
{Sp_{\beta}\hat\rho(t)\hat P'_{\beta j'}}
{Sp_{\alpha\beta}\hat\rho(t)\hat P'_{\beta j'}}\otimes\hat P_{\beta j'},
\label{8} \end{equation}
being the special case of the equation (3a).

Uniting (\ref{6}) $\equiv$ (\ref{1}) and (\ref{8}), it is possible
to formulate the generalized algorithm
of reduction as the procedure of calculating the density operator
 of an observable subsystem in the approximation of a given state
of the unobservable subsystem in the form of two consecutive operations:
\begin{enumerate}
\item {\bf A model of interaction of a
 closed quantum system with a measuring device is postulated from
physical reasons and is expressed mathematically in the form of the
approximation of a given state of
 an unobservable subsystem.}
\item {\bf The density operator of the observable subsystem
is calculated from equation (\ref{3a}) of correlated quasi-uncoupled subsystems
of a closed quantum
system.}
\end{enumerate}

In (\ref{6}) and (\ref{8}) two special approximations of the given state
of the subsystem $\beta$ are realized. In (\ref{6}) the subsystem $\beta$ is
permanently in each of eigenstates with equal probability
 and in (\ref{8}) it is with certainty  in one of them at the time $t$.

{\bf Consequence 3}

It is known \cite{fano,blum} that calculating the mean value
of the observable of a closed system
described by the operator $\hat A'$, the von Neumann's
 reduction of a closed system is carried out implicitly:
\begin{equation}
\langle\hat A'(t)\rangle=Sp_{\alpha\beta}\hat\rho(t)\hat A'=
Sp\hat\rho_{\alpha N}(t)\hat A=\langle\hat A(t)\rangle_{N}.\label{9}
\end{equation}
In doing so not the state of the
subsystem $\beta$ is postulated but the operator
of the closed system that corresponds to the observable
 $\hat A$ of the independent
subsystem $\alpha$ ($\hat A'\leftrightarrow\hat A$)
 (see \cite[Ch. 6.2]{von}). Hence, by Theorem and Consequence 2, the
choice of the required operator as $\hat A'$
is equivalent to realization of both approximations {\bf i)} and {\bf ii)}.

Such equivalence becomes evident by considering
the correlator of observables
 $\hat A$ and $\hat B$ of subsystems
 $\left\langle\left(\hat A\otimes\hat B\right)(t)\right\rangle=
Sp_{\alpha\beta}\hat\rho(t)\left(\hat A\otimes\hat B
\right)$.
If we use the obvious identity
$\hat A\otimes\hat B=
\left(\hat A\otimes\hat 1_{\beta}\right)
\left(\hat 1_{\alpha}\otimes\hat B\right)$,
the commutativity
 $\left[\hat A',\hat B'\right]=0$
and the interchangeability of partial traces
$ Sp_{\alpha\beta}=Sp_{\beta\alpha}$,
the correlator can be presented in two forms symmetric relative to
subsystems:
\begin{equation}
\left\langle\left(\hat A\otimes\hat B\right)(t)\right\rangle=
Sp_{\alpha\beta}\rho(t)\left(\hat A\otimes\hat B
\right)=\left\{
\begin{array}{c}
\left\langle\hat A(t)\right\rangle_{B}
\left\langle\hat B(t)\right\rangle_{N},\\
\left\langle\hat A(t)\right\rangle_{N}
\left\langle\hat B(t)\right\rangle_{A},
\end{array}
\right.\label{10}
\end{equation}
where
\begin{equation}
\left\langle\hat A(t)\right\rangle_{B}=
Sp_{\alpha}\hat\rho_{\alpha B}(t)
\hat A,\;\;
\left\langle\hat B(t)\right\rangle_{A}=
Sp_{\beta}\hat\rho_{\beta A}(t)
\hat B,\label{11}
\end{equation}
\begin{equation}
\hat\rho_{\alpha B}(t)=
\frac{Sp_{\beta}\hat\rho(t)\hat\rho'_{\beta B}}
{ Sp_{\alpha\beta}\hat\rho(t)\hat\rho'_{\beta B}},\;\;
\hat\rho_{\beta A}(t)=
\frac{Sp_{\alpha}\hat\rho(t)\hat\rho'_{\alpha A}}
{Sp_{\alpha\beta}\hat\rho(t)\hat\rho'_{\alpha A}}.\label{12}
\end{equation}
\begin{equation}
\hat\rho_{\alpha A}=\hat A/Sp_{\alpha}\hat A,\;\;
\hat\rho_{\beta B}=\hat B/Sp_{\beta}\hat B,\label{13}
\end{equation}
It is supposed here for simplicity that
 $\left\langle\hat A(t)\right\rangle_{N}\ne 0,
\left\langle\hat B(t)\right\rangle_{N}\ne 0,$ and $\hat A,\hat B$ are the
nonnegative operators.

The expressions (2.10)-(2.13) are the exact equations for arbitrary Hermitian
$\hat A,\hat B$.
The meaning of values $\left\langle\hat A(t)\right\rangle_{B}$ and
$\left\langle\hat B(t)\right\rangle_{A}$ is clear from the
comparison of (\ref{12}) with (\ref{3a}),(\ref{3b}). For example, the operator
$\hat\rho_{\alpha B}(t)$ is
the reduced density operator of the  subsystem $\alpha $ in the
approximation of the given state
$\hat\rho_{\beta B}$ of the subsystem $\beta$
in accordance with Consequence 2.
Accordingly,
$\left\langle\hat A(t)\right\rangle_{B}$ is
the mean value of the observable $\hat A$ of the subsystem
$\alpha$ if the subsystem $\beta$ is in the state $\hat\rho_{\beta B}$.
Thus, (\ref{10}) shows that the correlator
 $\left\langle\left(\hat A\otimes\hat B\right)(t)\right\rangle$
of the closed system $\hat H$ is equal to the product of mean values of
observables of subsystems calculated for
 the system $\hat H$ reduced in the different ways.
 Namely: 1)~the reduced state of a subsystem $\alpha$
is defined at the given $\hat\rho_{\beta}=\hat\rho_{\beta B} $, but
the reduced state of the subsystem $\beta $ is defined simultaneously at the
 given $\hat\rho_{\alpha}=\hat\rho_{\alpha min}$; or,
 leading to the same result,  2) the reduced state of the subsystem
$\alpha $
is defined at the given $\hat\rho_{\beta}=\hat\rho_{\beta min}$
but that of the
subsystem $\beta $
at $\hat\rho_{\alpha}=\hat\rho_{\alpha A}$.

In the specific case $\hat B=\hat 1_{\beta}$ we have
 $\hat\rho_{\beta B}=\hat\rho_{\beta min}, $
 $\left\langle\hat B(t)\right\rangle_{N}=
\left\langle\hat B(t)\right\rangle_{A}=1, $
 $\left\langle\hat A(t)\right\rangle_{B}=
\left\langle\hat A(t)\right\rangle_{N}$ and (\ref{10}) = (\ref{9}).

For this axiomatic choice of the operator of an observable
  of the closed system $\hat A'\leftrightarrow\hat A$,
  commonly accepted in the quantum mechanics,
 in the limits of uncoupled subsystems
($\hat H_{\alpha\beta}=0 $)
 the expression of the observable
$\langle\hat A'(t)\rangle=\langle\hat A(t)\rangle$ is exact.
But at $\hat H_{\alpha\beta}\ne 0$ the mathematically exact
equality (\ref{9})
 is the physical approximation as it  contains
 implicitly approximations {\bf i)} and {\bf ii)}.
One can see from (\ref{6}), (\ref{8}) - (\ref{13}) and Consequence 2 that
depending on a choice of a model of measurements (or of an algorithm of
reduction) the calculated mean value of the observable
 described by the same
 operator $\hat A$ and the same density operator of the closed system
 $\hat\rho(t)$ is different:
 $\langle\hat A(t)\rangle_{N}\ne\langle\hat A(t)\rangle_{Pj'}\ne
\langle\hat A(t)\rangle_{B}. $ The physical reason of this
difference is obvious --- a choice of a model of measurement (of a algorithm
of reduction)
 corresponds to a choice of the back effect of the measuring
 device on the state of the observable subsystem and, hence, on the result
of measurement. Or, alternatively, it is possible to say following
(\ref{10}) that each choice of the
model of measurement corresponds to the calculation of different
observables of the closed quantum system.

\section{ALGORITHM OF THE CORRELATED REDUCTION}\label{sec3}

If the approximation of the closed system is true, it seems reasonable to
seek the algorithm of reduction such that minimize the changing of
$\hat\rho(t)$.

Let us consider expressions:
\begin{equation}
\left\langle\hat A(t)\right\rangle_{C}=
Sp\hat\rho_{\alpha C}(t)\hat A,\qquad
\left\langle\hat B(t)\right\rangle_{C}=
Sp\hat\rho_{\beta C}(t)\hat B;\label{14}
\end{equation}
\begin{equation}
\left\langle\left(\hat A\otimes\hat B\right)(t)\right\rangle_{C}=
\left[Sp\hat\rho_{\alpha C}(t)\hat A\right]
\left[Sp\hat\rho_{\beta C}(t)\hat B\right].\label{15}
\end{equation}
The equalities (\ref{14}) define mean values of observables $\hat A$ and $
\hat B$
of independent subsystems with the correlated density operators
$\hat\rho_{\alpha C}(t)$ and $\hat\rho_{\beta C}(t)$, respectively.
But these correlated density operators of subsystems can be
determined only approximately for the lack of the exact expression of
$\hat\rho(t)$ in the form of (\ref{2}) at $\hat H_{\alpha\beta}\ne 0$. Let us
calculate $\hat\rho_{\alpha C}(t),\hat\rho_{\beta C}(t)$ from the
system of equations (\ref{3a}),(\ref{3b}) by the successive
approximation method \cite{solo,soloann}:
\begin{equation}
\hat\rho_{\alpha C}(t)=\lim_{n\to\infty}
\hat\rho_{\alpha}^{(n+1)}(t)=\lim_{n\to\infty}\frac{Sp_{\beta}
\hat\rho(t)\hat\rho_{\beta}^{(n)}(t)}
{Sp_{\alpha\beta}\hat\rho(t)\hat\rho_{\beta}^{(n)}(t)},\label{16a}
\end{equation}
\begin{equation}
\hat\rho_{\beta C}(t)=\lim_{n\to\infty}
\hat\rho_{\beta}^{(n+1)}(t)=\lim_{n\to\infty}\frac{Sp_{\alpha}
\hat\rho(t)\hat\rho_{\alpha}^{(n)}(t)}
{Sp_{\alpha\beta}\hat\rho(t)\hat\rho_{\alpha}^{(n)}(t)},\label{16b}
\end{equation}
where, for an example, the von Neumann's reduced density operators
$\hat\rho_{\alpha N}(t),\hat\rho_{\beta N}(t)$  may be used as a
zeroth-order (n=0) iteration.
For the convergent sequences (\ref{16a}),(\ref{16b}), the calculated
self-congruent
density operators of subsystems most closely correspond to (\ref{2}). It
 means that the mathematical operation of reduction (\ref{16a}),(\ref{16b})
 does not
change approximately the state of the closed quantum system and hence may be
considered to correspond to the quantum nondemolition measurement.

A matter of principle is if (\ref{9}) or (\ref{14}) describes observables of
subsystems of a closed system more properly.

The expression (\ref{9}) is the mathematical equality but the physical
approximation by Consequences 1-3. Each of (\ref{14}) is the physical
equality but mathematical approximation by (\ref{2}) and (\ref{16a}),
(\ref{16b}).
The accuracy of (\ref{14}) or (\ref{16a}),(\ref{16b}) can be estimated by
a comparison of
matrix elements at the left-hand and right-hand sides of (\ref{2}) or by
a comparison
 of mean values at the left-hand and right-hand sides of (\ref{15}).
It should be particularly emphasized that the correlated reduction
(\ref{16a}),(\ref{16b})
seems to be more appropriate to the equivalency of subsystems of
the closed system than the reduction (\ref{1}) including the physically
impossible property ($T=\infty$) of the unobservable subsystem. Therefore
we expect that the correlated self-congruent density operators (\ref{16a}),
(\ref{16b})
would more properly describe the dynamics of subsystems in experimental cases
when the approximation of the closed system holds.

Let us consider the previous procedure of a correlated reduction and its
physical consequences on model problems of quantum optics.\\

\section{DYNAMICS OF A PAIR OF CORRELATED TWO-LEVEL SUBSYSTEMS}\label{sec4}

Two coupled two-level
subsystems constitute the compound  closed quantum system now in use
to describe many basic problems of
quantum optics and quantum informatics \cite{klysh,men,kilin}. Let us look
after the
dynamics of such simplest correlated subsystems. To visualize the
distinction in the dynamics of subsystems described by the
traditional reduction (\ref{1}),(\ref{6}), and by the
correlated reduction (\ref{3a}),(\ref{3b}),(\ref{16a}),(\ref{16b}) we
present at first the consideration in general.

Let us designate eigenstates of uncoupled subsystems
 $\alpha$ and $\beta$ by
 $\vert 2\rangle_{\alpha},\vert 1\rangle_{\alpha}$ and
$\vert 2\rangle_{\beta},\vert 1\rangle_{\beta}$, respectively.
Operators of the closed system are represented by matrices of the fourth
order, each element of which is designated by two pairs of
subscripts (see, e.g., \cite[Appendix A]{blum}). The first subscript of each
pair is attributed to a subsystem
 $\alpha$ and second subsctipt to a subsystem
$\beta$. The general expression of the density matrix
of the considered closed system is then:
\begin{equation}
\left(\hat\rho\right)=\left(
\begin{array}{cccc}
\rho_{22,22} & \rho_{22,21} & \rho_{22,12} & \rho_{22,11} \\
\rho_{21,22} & \rho_{21,21} & \rho_{21,12} & \rho_{21,11} \\
\rho_{12,22} & \rho_{12,21} & \rho_{12,12} & \rho_{12,11} \\
\rho_{11,22} & \rho_{11,21} & \rho_{11,12} & \rho_{11,11} \\
\end{array}
\right).\label{17}
\end{equation}
For an observable subsystem $\alpha$ the reduced von Neumann's
density matrix in accordance with (\ref{1}) is defined by expression:
\begin{equation}
\left(\hat\rho_{\alpha N}\right)=
\left(
\begin{array}{cc}
\rho_{22,22}+\rho_{21,21} & \rho_{22,12}+\rho_{21,11} \\
\rho_{12,22}+\rho_{11,21} & \rho_{12,12}+\rho_{11,11}
\end{array}
\right).\label{18}
\end{equation}
Accordingly, for an observable subsystem $\beta$:
\begin{equation}
\left(\hat\rho_{\beta N}\right)=
\left(
\begin{array}{cc}
\rho_{22,22}+\rho_{12,12} & \rho_{22,21}+\rho_{12,11}\\
\rho_{21,22}+\rho_{11,12} & \rho_{21,21}+\rho_{11,11}
\end{array}
\right).\label{19}
\end{equation}
In accordance with (\ref{6}), such definition of
the operation of reduction
is equivalent to the replacement
of (\ref{17}) by expressions:
\begin{eqnarray}
\left(\hat\rho\right)\to&&\left(\hat\rho_{R\alpha N}\right)=
\left(\hat\rho_{\alpha N}\right)\otimes\left(\hat\rho_{\beta\,min}\right)
\nonumber\\
&&=\frac12
\left(
\begin{array}{cccc}
\rho_{22,22}+\rho_{21,21} & 0 & \rho_{22,12} + \rho_{21,11} & 0 \\
 0 & \rho_{22,22}+\rho_{21,21} & 0 & \rho_{22,12}+\rho_{21,11} \\
\rho_{12,22}+\rho_{11,21} & 0 & \rho_{12,12}+\rho_{11,11} & 0 \\
 0 & \rho_{12,22}+\rho_{11,21} & 0 & \rho_{12,12}+\rho_{11,11}
\end{array}
\right),\label{20}
\end{eqnarray}
or
\begin{eqnarray}
\left(\hat\rho\right)\to&&\left(\hat\rho_{R\beta N}\right)=
\left(\hat\rho_{\alpha\, min}\right)\otimes\left(\hat\rho_{\beta N}\right)
\nonumber\\
&&=\frac12
\left(
\begin{array}{cccc}
\rho_{22,22}+\rho_{12,12} & \rho_{22,21}+\rho_{12,11} & 0 & 0 \\
\rho_{21,22}+\rho_{11,12} & \rho_{21,21}+\rho_{11,11} & 0 & 0 \\
 0 & 0 & \rho_{22,22}+\rho_{12,12} & \rho_{22,21}+\rho_{12,11} \\
 0 & 0 & \rho_{21,22}+\rho_{11,12} & \rho_{21,21}+\rho_{11,11} \\
\end{array}
\right).\label{21}
\end{eqnarray}
For comparison, we write the system of equations (\ref{3a}),(\ref{3b})
determining correlated reduced density matrices of subsystems
of the closed system (\ref{17}):
\begin{eqnarray}
\left(\hat\rho_{\alpha C}\right)=
\left(
\begin{array}{cc}
\alpha_{22} & \alpha_{21} \\
\alpha_{12} & \alpha_{11}
\end{array}
\right)\simeq&&
\left(
\begin{array}{cc}
\begin{array}{c}
(\rho_{22,22}\beta_{22}+\rho_{21,21}\beta_{11}+\\
+ \rho_{22,21}\beta_{12}+\rho_{21,22}\beta_{21})
\end{array}
\begin{array}{c}
(\rho_{22,12}\beta_{22}+\rho_{21,11}\beta_{11}+\\
+\rho_{22,11}\beta_{12}+\rho_{21,12}\beta_{21})
\end{array}
\\
\begin{array}{c}
(\rho_{12,22}\beta_{22}+\rho_{11,21}\beta_{11}+\\
+\rho_{12,21}\beta_{12}+\rho_{11,22}\beta_{21})
\end{array}
\begin{array}{c}
(\rho_{12,12}\beta_{22}+\rho_{11,11}\beta_{11}+\\
+\rho_{12,11}\beta_{12}+\rho_{11,12}\beta_{21})
\end{array}
\end{array}
\right)\nonumber\\
&&\times
\left(\rho_{22,22}\beta_{22}+\rho_{21,21}\beta_{11}+
\rho_{22,21}\beta_{12}+
\rho_{21,22}\beta_{21}\right.\nonumber\\
&&\left.
+\rho_{12,12}\beta_{22}+\rho_{11,11}\beta_{11}+
\rho_{12,11}\beta_{12}+\rho_{11,12}\beta_{21}
\right)^{-1},\label{22}
\end{eqnarray}
\begin{eqnarray}
\left(\hat\rho_{\beta C}\right)=
\left(
\begin{array}{cc}
\beta_{22} & \beta_{21} \\
\beta_{12} & \beta_{11}
\end{array}
\right)\simeq&&
\left(
\begin{array}{cc}
\begin{array}{c}
(\rho_{22,22}\alpha_{22} + \rho_{12,12}\alpha_{11}+ \\
+\rho_{22,12}\alpha_{12} + \rho_{12,22}\alpha_{21})
\end{array}
&
\begin{array}{c}
(\rho_{22,21}\alpha_{22} + \rho_{12,11}\alpha_{11}+ \\
+\rho_{22,11}\alpha_{12} + \rho_{12,21}\alpha_{21})
\end{array}
\\
\begin{array}{c}
(\rho_{21,22}\alpha_{22} + \rho_{11,12}\alpha_{11}+ \\
+\rho_{21,12}\alpha_{12} + \rho_{11,22}\alpha_{21})
\end{array}
&
\begin{array}{c}
(\rho_{21,22}\alpha_{22} + \rho_{11,11}\alpha_{11}+ \\
+\rho_{21,11}\alpha_{12} + \rho_{11,21}\alpha_{21})
\end{array}
\end{array}
\right)\nonumber\\
&&\times
\left(\rho_{22,22}\alpha_{22}+ \rho_{12,12}\alpha_{11}+
\rho_{22,12}\alpha_{12}+
\rho_{12,22}\alpha_{21}\right.\nonumber\\
&&+\left. \rho_{21,21}\alpha_{22}+ \rho_{11,11}\alpha_{11}+
\rho_{21,11}\alpha_{12}+ \rho_{11,21}\alpha_{21}
\right)^{-1}.\label{23}
\end{eqnarray}
It is seen that in the general case
the mutual correlation of subsystems (\ref{22}) and (\ref{23}) as distinct
from (\ref{18}) and (\ref{19}) implies  the following.\\
1) The possibility of the simultaneous existence of quantum coherence, i.e.
 nonzero nondiagonal elements of reduced density matrices
 of both subsystems
$(\alpha_{12}\ne 0, \beta_{12}\ne 0) $. \\
2) The inequality of probabilities to find the subsystems in
their eigenstates $(\alpha_{11}\ne\alpha_{22},
 \beta_{11}\ne\beta_{22}) $.

 Let us use (\ref{17})-(\ref{23}) in considering two known physical problems.

\subsection{Representation of an entangled EPR (Einstein-Podolsky-Rosen) pair
of photons as the superposition of separated
correlated photons}\label{subsec4.1}

The model of the entangled EPR state is of considerable current use in
the description of experiments on the correlation of photons
\cite{klysh,men,kilin}.
The density matrix of such closed quantum system of two noninteracting
but correlated photons in designations (\ref{17}) is of the form
\cite{klysh,men,kilin}:
\begin{equation}
\left(\hat\rho_{S}\right)
=\frac12
\left(
\begin{array}{cccc}
0 & 0 & 0 & 0 \\
0 & 1 & -1 & 0  \\
0 & -1 & 1 & 0 \\
0 & 0 & 0 & 0 \\
\end{array}
\right).\label{24}
\end{equation}
States of subsystems (i.e. individual photons) by the von Neumann's reduction
(\ref{1}),(\ref{18}), (\ref{19}) are:
\begin{equation}
\left(\hat\rho_{\alpha N}\right)=\left(\hat\rho_{\alpha\, min}\right)=
\frac12
\left(
\begin{array}{cc}
1 & 0 \\
0 & 1
\end{array}
\right),\;
\mbox{or}\;
\left(\hat\rho_{\beta N}\right)=\left(\hat\rho_{\beta\, min}\right)=
\frac12
\left(
\begin{array}{cc}
1 & 0 \\
0 & 1
\end{array}
\right).\label{25}
\end{equation}
By (\ref{6}),(\ref{20}),(\ref{21}),  for both subsystems it is equivalent
to the replacement
of the density matrix (\ref{24}) by the reduced density matrix
\begin{equation}
\left(\hat\rho_{S} \right)\to
 \left(\hat\rho_{SRN} \right)=\hat\rho_{min}=
 \frac14 \hat 1.\label{26}
\end{equation}
The expressions (\ref{25}) and (\ref{26}) show that each photon can
 be registered in any eigenstate with equal probability (1/2) but
 they do not
determine the mutual correlation of results of simultaneous or
consecutive measurements of both photons. The approximate
nature of the von Neumann's reduction makes itself evident in the
nonequality  (\ref{24})$\ne$(\ref{26}).
On the other hand, as photons in the pure state (\ref{24}) are noninteracting,
one might expect the exact expression of the system (\ref{24}) through
correlated subsystems in terms similar to (\ref{16a}),(\ref{16b}) to exist.
With (\ref{22}) and (\ref{23}) for
 (\ref{24}) the iterative procedure (\ref{16a}),(\ref{16b}) converges
 already in the
first iteration for the arbitrary zeroth-order iteration and
results in the following relations between the elements of correlated
reduced density matrices of subsystems:
$\alpha_{22}=\beta_{11},\,\alpha_{21}=-\beta_{21}$, in addition to usual
relations of normalizing and of hermiticity: $\alpha_{11}=1-\alpha_{22},\,
\alpha_{12}=\alpha^{*}_{21}$. With these correlation conditions,
let us write down the system of equations for calculating
the elements
$\alpha_{22},\,\alpha_{21}$ from (\ref{2}).
This system is separated into three groups of equations:
for diagonal elements,
nondiagonal elements and equations of their relation:
\begin{equation}
\begin{array}{lll}
\alpha_{22}\beta_{22}\leftrightarrow 0=:\alpha_{22}(1-\alpha_{22}), &
\alpha_{21}\beta_{12}\leftrightarrow -1/2=:-\vert\alpha_{21}\vert^2, &
\alpha_{22}\beta_{21}\leftrightarrow 0=:\alpha_{22}(-\alpha_{21}), \\
\alpha_{22}\beta_{11}\leftrightarrow 1/2=:\alpha_{22}^2, &
\alpha_{21}\beta_{21}\leftrightarrow 0=:-(\alpha_{21})^2, &
\alpha_{21}\beta_{22}\leftrightarrow 0=:\alpha_{21}(1-\alpha_{22}), \\
\alpha_{11}\beta_{22}\leftrightarrow 1/2=:(1-\alpha_{22})^2, &  &
\alpha_{21}\beta_{11}\leftrightarrow 0=:\alpha_{21}\alpha_{22}, \\
\alpha_{11}\beta_{11}\leftrightarrow 0=:(1-\alpha_{22})\alpha_{22}, &  &
\alpha_{11}\beta_{21}\leftrightarrow 0=:-(1-\alpha_{22})\alpha_{21}. \\
\end{array}\label{27}
\end{equation}
The system (\ref{27}) is obviously incompatible as an algebraic system of
equation,
but it is a compatible system of statistical equations that we have
designated
by  "$ =: $" instead of "$ = $".
For example, the system of four equations from the first column of (\ref{27})
is compatible, if it is granted that $\alpha_{22}$ takes
two values:
$\alpha^{(1)}_{22}=1$ and $\alpha^{(2)}_{22}=0 $
with equal probability (1/2), and each equation
is considered as a result of statistical averaging.
In this way, for the first equation of the first column from (\ref{27})
we have:
$0=(1/2)\left[\alpha^{(1)}_{22}(1-\alpha^{(1)}_{22})+
\alpha^{(2)}_{22}(1-\alpha^{(2)}_{22})\right].$
For the second equation:
$1/2=(1/2)\left[\left(\alpha^{(1)}\right)^2+
\left(\alpha^{(2)}\right)^2\right]$.
In solving the system (\ref{27}) as a system of the statistical
equations, we obtain within the arbitrary phase factor ($\exp(i\theta)$):
\begin{eqnarray}
\left(\hat\rho_{S}\right)=
\sum_{i=1}^{4}p_i\left\{\left(\hat\rho_{\alpha C}^{(i)}\right)\otimes
\left(\hat\rho_{\beta C}^{(i)}\right)\right\}\qquad\qquad\qquad\nonumber \\
=\frac14\left\{
\left(
\begin{array}{cc}
1 & \frac{1}{\sqrt{2}}e^{i\theta} \\
\frac{1}{\sqrt{2}}e^{-i\theta} & 0
\end{array}
\right)\otimes
\left(
\begin{array}{cc}
0 & \frac{1}{\sqrt{2}}e^{i(\theta+\pi)} \\
\frac{1}{\sqrt{2}}e^{-i(\theta+\pi)} & 1
\end{array}
\right)
\right.\qquad\nonumber\\
+\left(
\begin{array}{cc}
1 & \frac{1}{\sqrt{2}}e^{i(\theta+\pi)} \\
\frac{1}{\sqrt{2}}e^{-i(\theta+\pi)} & 0
\end{array}
\right)\otimes
\left(
\begin{array}{cc}
0 & \frac{1}{\sqrt{2}}e^{i\theta} \\
\frac{1}{\sqrt{2}}e^{-i\theta} & 1
\end{array}
\right)\qquad\nonumber\\
+\left(
\begin{array}{cc}
0 & \frac{1}{\sqrt{2}}e^{i(\theta+\pi/2)} \\
\frac{1}{\sqrt{2}}e^{-i(\theta+\pi/2)} & 1
\end{array}
\right)\otimes
\left(
\begin{array}{cc}
1 & \frac{1}{\sqrt{2}}e^{i(\theta+3\pi/2)} \\
\frac{1}{\sqrt{2}}e^{-i(\theta+3\pi/2)} & 0
\end{array}
\right)\nonumber\\
+
\left.\left(
\begin{array}{cc}
0 & \frac{1}{\sqrt{2}}e^{i(\theta+3\pi/2)} \\
\frac{1}{\sqrt{2}}e^{-i(\theta+3\pi/2)} & 1
\end{array}
\right)\otimes
\left(
\begin{array}{cc}
1 & \frac{1}{\sqrt{2}}e^{i(\theta+\pi/2)} \\
\frac{1}{\sqrt{2}}e^{-i(\theta+\pi/2)} & 0
\end{array}
\right)
\right\}.\label{28}
\end{eqnarray}
Consequently, the entangled pure EPR state is exactly represented as a sum
of four hidden equiprobable ($p_i=const=1/4$) states of separated but
correlated ($\hat\rho_{\alpha C}^{(i)},\hat\rho_{\beta C}^{(i)}$) subsystems.
Each of summands presents by itself the distinct state of EPR pare. Each
of $\hat\rho_{\alpha C}^{(i)},\hat\rho_{\beta C}^{(i)}$ corresponds to the
existence of the subsystem in one of eigenstates: the diagonal elements of
density matrices are 1 and 0 only. Note simultaneously the unusual nonzero
values of nondiagonal elements. Nondiagonal elements of this type reflect not
the usual superposition of eigenstates of one subsystem but the correlation in
phase of the eigenstates of the correlated subsystems.

The considered model of a pair of noninteracting two-level
subsystems is the elementary example of the compound closed quantum system.
The obvious next step of the generalization of the model is
taking into account the interaction between subsystems.

\subsection{Dynamics of a pair of interacting two-level atoms}
\label{subsec4.2}

The interaction of a pair of two-level subsystems was in detail considered
in the theory
of magnetic resonance, therefore it is convenient to take advantage of
designations accepted there. The hamiltonian of a pair of identical
spins-1/2 in an external d-c magnetic field
with a simple form of the spin-spin interaction is
\cite{abragam}:
\begin{equation}
\left(\hat H\right)
=\hbar\left(
\begin{array}{cccc}
\omega+J &  0 &  0 &     d     \\
0        & -J &  c &     0     \\
0        &  c & -J &     0     \\
d        &  0 &  0 & -\omega+J \\
\end{array}
\right),\label{29}
\end{equation}
where $\hbar\omega$ is the Zeeman energy of a
 single spin-1/2 in an external d-c
magnetic field, parameters $J, c, d $ characterize the energy of the
spin-spin interaction, which for definiteness is believed to be
rather weak $(\omega > J, c, d)$.
Designations of basic states
accepted in (\ref{17}) are used, i.e. $\vert 2\rangle, \vert 1\rangle $
are eigenstates of a single spin-1/2 in the d-c magnetic field with higher
and lower energies, respectively.

The operator of evolution of quantum system (\ref{29}) results in:
\begin{eqnarray}
&&\exp\left(\mp\frac{it}{\hbar}\hat H\right)=
\left\{
\begin{array}{c}
\hat U(t)\\
\hat U^{\dagger}(t)
\end{array}
\right\}=\nonumber\\
&&
\left(
\begin{array}{cccc}
e^{\mp iJt}\left[C(\Omega t)\mp\frac{i\omega}{\Omega}S(\Omega t)\right]
 & 0 & 0 & \mp(id/\Omega)e^{\mp iJt}S(\Omega t) \\
0 & e^{\pm iJt}\cos(ct) & \mp ie^{\pm iJt}\sin(ct)  & 0  \\
0 & \mp ie^{\pm iJt}\sin(ct) & e^{\pm iJt}\cos(ct) & 0 \\
\mp(id/\Omega)e^{\mp iJt}S(\Omega t) & 0 & 0 &
e^{\mp iJt}\left[C(\Omega t)\pm\frac{i\omega}{\Omega}S(\Omega t)\right] \\
\end{array}
\right),
\label{30}
\end{eqnarray}
where $C(\Omega t)=\cos(\Omega t),S(\Omega t)=\sin(\Omega t)$.
Thus, the density matrix $ (\hat\rho (t)) $, i.e. the dynamics of the system
(\ref{29})
at the given initial condition $ (\hat\rho (0)) $ is known exactly.
To describe the dynamics of each spin one should calculate
correlated reduced
 density matrices of spin-1/2 subsystems
according to Sec. \ref{sec3}. Let the initial state of the system
(\ref{29}) be:
\begin{equation}
\left(\hat\rho(0)\right)
=
\left(
\begin{array}{cccc}
0 & 0 & 0 & 0 \\
0 & \cos^2\varphi & -\sin\varphi\cos\varphi & 0  \\
0 & -\sin\varphi\cos\varphi & \sin^2\varphi & 0 \\
0 & 0 & 0 & 0 \\
\end{array}
\right).
\label{31}
\end{equation}
For example, the initial condition
of a radical pair born during photosynthesis is described in this form
\cite{salikh}.
Similarly
to the consideration of Subsec. \ref{subsec4.1}, this initial density matrix
 is presented as:
\begin{eqnarray}
&&\hat\rho(0)\simeq
\hat\rho_{RC}(0)\nonumber\\
&&=
\left\{
\begin{array}{cc}
\cos^2\varphi\left(
\begin{array}{cc}
1 & 0 \\
0 & 0
\end{array}
\right)\otimes
\left(
\begin{array}{cc}
0 & 0 \\
0 & 1
\end{array}
\right)
+\sin^2\varphi\left(
\begin{array}{cc}
0 & 0 \\
0 & 1
\end{array}
\right)\otimes
\left(
\begin{array}{cc}
1 & 0 \\
0 & 0
\end{array}
\right),
& \vert\tan\varphi\vert>1,\\
\cos^2\varphi\left(
\begin{array}{cc}
0 & 0 \\
0 & 1
\end{array}
\right)\otimes
\left(
\begin{array}{cc}
1 & 0 \\
0 & 0
\end{array}
\right)
+\sin^2\varphi\left(
\begin{array}{cc}
1 & 0 \\
0 & 0
\end{array}
\right)\otimes
\left(
\begin{array}{cc}
0 & 0 \\
0 & 1
\end{array}
\right),
&\vert\tan\varphi\vert <1,\\
\hat\rho_{\mbox{\scriptsize S}}, & \tan\varphi=1,\\
\hat\rho_{\mbox{\scriptsize T}}, & \tan\varphi=-1,
\end{array}
\right.
,
\label{32}
\end{eqnarray}
where
\begin{eqnarray}
\left(\hat\rho_{\mbox{\scriptsize T}}\right)=&&\frac12
\left(
\begin{array}{cccc}
0 & 0 & 0 & 0 \\
0 & 1 & 1 & 0 \\
0 & 1 & 1 & 0 \\
0 & 0 & 0 & 0
\end{array}
\right)=
\frac14\left[
\left(
\begin{array}{cc}
1 & \frac{1}{\sqrt 2}e^{i\theta}\\
\frac{1}{\sqrt 2}e^{-i\theta} & 0
\end{array}
\right)
\otimes
\left(
\begin{array}{cc}
0 & \frac{1}{\sqrt 2}e^{i\theta}\\
\frac{1}{\sqrt 2}e^{-i\theta} & 1
\end{array}
\right)\right.\nonumber\\
&&+\left(
\begin{array}{cc}
0 & \frac{1}{\sqrt 2}e^{i(\theta+\pi/2)}\\
\frac{1}{\sqrt 2}e^{-i(\theta+\pi/2)} & 1
\end{array}
\right)
\otimes
\left(
\begin{array}{cc}
1 & \frac{1}{\sqrt 2}e^{i(\theta+\pi/2)}\\
\frac{1}{\sqrt 2}e^{-i(\theta+\pi/2)} & 0
\end{array}
\right)
\nonumber\\
&&+
\left(
\begin{array}{cc}
1 & \frac{1}{\sqrt 2}e^{i(\theta+\pi)}\\
\frac{1}{\sqrt 2}e^{-i(\theta+\pi)} & 0
\end{array}
\right)
\otimes
\left(
\begin{array}{cc}
0 & \frac{1}{\sqrt 2}e^{i(\theta+\pi)}\\
\frac{1}{\sqrt 2}e^{-i(\theta+\pi)} & 1
\end{array}
\right)\nonumber\\
&&+\left.
\left(
\begin{array}{cc}
0 & \frac{1}{\sqrt 2}e^{i(\theta+3\pi/2)}\\
\frac{1}{\sqrt 2}e^{-i(\theta+3\pi/2)} & 1
\end{array}
\right)
\otimes
\left(
\begin{array}{cc}
1 & \frac{1}{\sqrt 2}e^{i(\theta+3\pi/2)}\\
\frac{1}{\sqrt 2}e^{-i(\theta+3\pi/2)} & 0
\end{array}
\right)
\right].
\label{33}
\end{eqnarray}
Unlike (\ref{28}) expression (\ref{32}) is exact only for
$\vert\tan\varphi\vert=1 $ (i.e. for the last two rows in (\ref{32})).

As a result of correlated reduction (\ref{16a}),(\ref{16b}) the density matrix
 $ \hat\rho (t) $ is represented at any time as follows:
\begin{eqnarray}
\left(\hat\rho(t)\right)=
\frac12\left(
\begin{array}{cccc}
0 & 0 & 0 & 0 \\
0 & P(\varphi,t) & i\cos(2\varphi)\sin(2ct)-\sin(2\varphi) & 0  \\
0 & -i\cos(2\varphi)\sin(2ct)-\sin(2\varphi) & M(\varphi,t) & 0 \\
0 & 0 & 0 & 0 \\
\end{array}
\right)
\to (\hat\rho_{RC}(t)\nonumber\\
=
\left\{
\begin{array}{cc}
\frac12\left[P(\varphi,t)
\left(
\begin{array}{cc}
1 & 0 \\
0 & 0
\end{array}
\right)\otimes
\left(
\begin{array}{cc}
0 & 0 \\
0 & 1
\end{array}
\right)
+M(\varphi,t)
\left(
\begin{array}{cc}
0 & 0 \\
0 & 1
\end{array}
\right)\otimes
\left(
\begin{array}{cc}
1 & 0 \\
0 & 0
\end{array}
\right)\right],
& C(\varphi,t)>0,\\
\frac12\left[P(\varphi,t)
\left(
\begin{array}{cc}
0 & 0 \\
0 & 1
\end{array}
\right)\otimes
\left(
\begin{array}{cc}
1 & 0 \\
0 & 0
\end{array}
\right)
+M(\varphi,t)
\left(
\begin{array}{cc}
1 & 0 \\
0 & 0
\end{array}
\right)\otimes
\left(
\begin{array}{cc}
0 & 0 \\
0 & 1
\end{array}
\right)\right],
& C(\varphi,t)<0,\\
\hat\rho_{\mbox{\scriptsize S}}, & \tan(\varphi)=1,\\
\hat\rho_{\mbox{\scriptsize T}}, & \tan(\varphi)=-1,\\
\end{array}
\right.
\label{34}
\end{eqnarray}
where $C(\varphi,t)=\cos(2\varphi)\cos(2ct),  \;P(\varphi,t)=1+
C(\varphi,t),\;
M(\varphi,t)=1-C(\varphi,t)$.

Thus, the dynamics of the closed system of two interacting  spins-1/2
(\ref{29}), according to the
initial state (the value of $\varphi$), may be of two types.
For $\vert\tan\varphi\vert=1$ the system is in the stable
statistically defined state
S$\leftrightarrow$(\ref{28}) or T$\leftrightarrow$(\ref{33}).
For $\vert\tan\varphi\vert\ne1$ there are the harmonic oscillations
of probabilities to detect spins in basic states opposite in phase.
At $t_n=(2n+1)\pi/8c,\,(n=0,1,2,\ldots)$ for an arbitrary initial state
these probabilities are the same and equal 1/2.

\section{DYNAMICS OF A TWO-LEVEL ATOM IN A RESONANT FIELD}\label{sec5}

To demonstrate the decoherence phenomenon,
the model problem of quantum optics
on the dynamics of the inversion of a two-level atom interacting
with a resonant electromagnetic field of a single-mode cavity
(see, e.g., \cite{men,haroche}) is used traditionally.
It is attractive due to the existence
of the experimental setup of the cavity quantum electrodynamics entangling
machine
\cite{haroche} and  the existence of the exact solution of the theoretical
Janes-Cummings model (JCM) \cite{louis,allen}.

Let us trace basic distinctions in the dynamics of atom and field
following from
 definitions (\ref{1})$\leftrightarrow$(\ref{6}) and
 (\ref{3a}),(\ref{3b})$\leftrightarrow$(\ref{16a}),(\ref{16b}).

The hamiltonian of the resonant JCM looks like
\cite{louis,allen,yoo,acker,optacta,demid}:
\begin{equation}
\hat H=\hat H_a+\hat H_f+\hat H_{af}=\frac{\hbar\omega}{2}\left(\hat P_{22}-
\hat P_{11}\right)+\frac{\hbar\omega}{2}\left(\hat a^+\hat a+\hat a\hat
a^+\right)+\frac{i\hbar\Omega}{2}\left(\hat P_{21}\hat a-\hat P_{12}\hat
a^+\right), \label{35}
\end{equation}
where $\hat H_a, \hat H_f, \hat H_{af} $ are hamiltonians of the atom, field
and atom-field interactions, $\omega$ is the resonant frequency of the field,
$\Omega$ is the constant of the atom-field interaction,
 $\hat P_{ij}$ are projection atomic operators
representing square matrices of the second order, the $ij$-th  element of
which is equal to 1, others are equal to 0, $i,j=1,2 $ and
the subscript 2 corresponds to the level with the  higher energy.

It is convenient to write down
the solution of the JCM (obtained in a general form in the Heisenberg
 picture in Ref. \cite{acker})
 in the Schr\"odinger picture in the form of
expression for the operator of evolution \cite{optacta,demid}:
\begin{eqnarray}
&&\left\{
\begin{array}{c}
\hat U(t)\\
\hat U^{\dagger}(t)
\end{array}
\right\}=\exp\left(\mp\frac{it}{\hbar}\hat
H\right)=\hat P_{22}\exp\left(\mp i\omega t\hat a\hat a^+\right)\,
\cos\left[\frac{\Omega t}{2}\left(\hat a\hat a^+\right)^{1/2}\right]
\nonumber\\
&&+\hat P_{11}\exp\left(\mp i\omega t\hat a^+\hat a\right)\,
\cos\left[\frac{\Omega t}{2}\left(\hat a^+\hat a\right)^{1/2}\right]\pm
\hat P_{21}\exp\left(\mp i\omega t\hat a\hat a^+\right)\,
\exp(i\hat\varphi)\,\sin\left[\frac{\Omega t}{2}\left(\hat a^+\hat
a\right)^{1/2}\right]\nonumber\\
&&\mp\hat P_{12}\exp\left(\mp i\omega t\hat a^+\hat a\right)\,
\exp(-i\hat\varphi)\,
\sin\left[\frac{\Omega t}{2}\left(\hat a\hat a^+\right)^{1/2}\right],
\label{36}
\end{eqnarray}
where $\exp(\pm i\hat\varphi)=\left\{\begin{array}{c}
\left(\hat a^+\hat a+1\right)^{-1/2}\hat a\\
\hat a^+\left(\hat a^+\hat a+1\right)^{-1/2}
\end{array}\right\}$ is the operator of a phase of a field \cite{loudon}.
 Use of the operators $\exp(\pm i\hat\varphi)$ and
 $\hat P_{ij}$
 (instead of more widespread Pauli operators) allows us to write down
 the operator
 of evolution in the symmetric compact form
 (\ref{36}), convenient for the further analysis.

The elementary example (traditionally used in
 educational purposes \cite{louis,allen}) is the consideration of the
 dynamics  of the inversion of the atom, if the initial condition was
  $\hat\rho(0)=\hat\rho_{a}(0)\hat\rho_{f}(0)=
\hat P_{22}\vert 0\rangle\langle 0\vert $, i.e. the atom was in the
excited state $\vert 2\rangle$ and the field was in the vacuum state.
The  known result, at once following from (\ref{36}) and (\ref{1}),
consists in oscillations with the frequency $\Omega$
 of the inversion  of the atom or the probability of the existence of
the photon in the cavity (vacuum Rubi oscillation \cite{haroche}):
\begin{equation}
\hat\rho_{aN}(t)=\hat P_{22}\cos^2(\Omega t/2)+\hat P_{11}\sin^2(\Omega t/2),
\label{37}
\end{equation}
or
\begin{equation}
\hat\rho_{fN}(t)=\sin^2(\Omega t/2)\vert 1\rangle\langle 1\vert +
\cos^2 (\Omega t/2) \vert 0\rangle\langle 0\vert.
\label{38}
\end{equation}
At $t\ne 2\pi l/\Omega, \; (l=0,1,2, \ldots) $ the atom and the
field are in a mixture state corresponding
to decoherence of the observed subsystem.

It is easy to notice that expressions (\ref{37}) and (\ref{38}) are
incompatible for the system (\ref{3a}),(\ref{3b}).
We search for the correlated reduced density operators
 of the atom and the field with (\ref{16a}),(\ref{16b}), using
 (\ref{37}) and (\ref{38}) as the zeroth-order iteration.
 As a result, we obtain:
\begin{equation}
\hat\rho_a(t)=C(t)\hat P_{22}+S(t)\hat P_{11},\quad
\hat\rho_f(t)=S(t)\vert 0\rangle\langle 0\vert+
C(t)\vert 1\rangle\langle 1\vert,  \label{39}
\end{equation}
where $S(t)=1-C(t),\; s=\sin(\Omega t/2),c=\cos(\Omega t/2)$,
\begin{equation}
C(t)=\lim_{n\to\infty}\frac{c^{2n}}{c^{2n}+s^{2n}}=\left\{
\begin{array}{cc}
1,       & [(4l-1)\pi]/2\Omega<t<[(4l+1)\pi]/2\Omega,\\
1/2, & t=[(2l+1)\pi]/2\Omega,\\
0,       & [(4l+1)\pi]/2\Omega<t<[(4l+3)\pi]/2\Omega.
\end{array}\right.
\label{40}
\end{equation}
According to (\ref{39}) and (\ref{40}), probabilities to find the atom
and the field
in their eigenstates are stepwise periodic functions of time
with the period $\pi/\Omega$.
 Practically all time the atom and the field are in pure states
and only at the discrete moments of time they are in a mixture state
equiprobable for two probable pure states.

The expression (\ref{37}) is employed at once to describe the experimentally
observed vacuum Rubi oscillations \cite{haroche,brune}. We call attention to
the experimental "dead time" prior to the first-time photon emission in these
experiments. The authors of \cite{haroche,brune} included the effective
interaction time taking
into account the spatial variation of the atom-field coupling when atom moves
across the cavity. It seems us unreasonable for the initially vacuum field
up to the first photon emission. All the space: the interior-space and the
outer-space of the cavity, is uniform for the vacuum field. Thus all the
atom way after the excitation must be included into consideration and not
just the interior of the cavity. Then the experimental results
\cite{haroche,brune} contain the dead time when atom is in the excited state
and there is not spontaneous emission that corresponds qualitatively to the
time interval $(0\to\pi/2\Omega)$ of (\ref{39}),(\ref{40}).

The other experimental result corresponding qualitatively to
(\ref{39}),(\ref{40})  consist
in the observing of the omnidirectional superfluorescense
\cite{hartmann,lvovsky}. There is the well-known
experimental situation, when each atom of the system of many
uncoupled two-level
atoms for a short (in comparison with $\pi/\Omega$) time interval is
transferred into the
 excited state and the spontaneous emission of the inverted
system of atoms is investigated without cavity.
The pulse of radiation
 experimentally observed in this case after a time interval $\tau$
is traditionally
 interpreted, after the authors of first observation \cite{skrib},
as a pulse of superradiation or superfluorescence \cite{dicke}. And $\tau$ is
considered to be defined by the time of self-organization  in the system
of atoms at the expense of their interaction through a common
radiation field \cite{skrib}.

According to (\ref{40}), each atom emits a photon at the same time
 $\tau =\pi/2\Omega $ if at $t=0$ each atom was in the excited state,
 i.e. the pulsed emission of spontaneous radiation
by the  system of noninteracting atoms may be due to
the correlation of each atom with the initially vacuum field. And for the
gaseous spherical sample \cite{hartmann,lvovsky} the directions of the
spontaneous emission of photons are equiprobable.

\section{CONCLUSION}

Results of the present work may be formulated as
the following assertions:

\begin{enumerate}

\item The identification of an operator of an observable of a subsystem
of a closed quantum system with the appropriate extended operator
of the independent observable subsystem ($\hat A'\leftrightarrow\hat A$)
corresponds to the physical approximation,
at which this obserable is calculated provided that
the unobservable subsystem is in the steady state of minimum
 information (infinite temperature).

\item The known algorithms of reduction correspond to the approximation of
a given state of the unobservable subsystem.

\item The algorithm of correlated reduction that most corresponds to
 the ideal quantum measurement of observables of subsystems of the closed
quantum sistem may be
 presented as the self-congruent calculation of mutually correlated
reduced density operators of subsystems
$\hat\rho_{\alpha C}(t),\hat\rho_{\beta C}(t)$
by the successive
approximation method (\ref{16a}),(\ref{16b}). The mean values
of observebles of subsystems
at the ideal quantum measurements are defined by expressions
$\left\langle\hat A(t)\right\rangle_{C}=
Sp\hat\rho_{\alpha C}(t)\hat A,\,
\left\langle\hat B(t)\right\rangle_{C}=
Sp\hat\rho_{\beta C}(t)\hat B.$

\item The correlations in the dynamics of subsystems are demonstrated
with examples of simple compound quantum systems: 1) the pair of the coupled
two-level atoms (spins-1/2), and 2) the two-level atom in a single-mode
 resonant field.
\end{enumerate}

\begin{acknowledgments}
The author acknoledge numerous discussions with V. N. Lisin.
\end{acknowledgments}

\end{document}